\documentclass[aps,prl,onecolumn]{revtex4}
\usepackage{graphicx}
\usepackage{amsmath}
\usepackage[latin1]{inputenc} 
\begin{document}

\title{Determining the Minimum Uncertainty State of Nonclassical Light}

\author{C. Gabriel\footnote{ now at: Max Planck Institute for the Science of Light, Günther-Scharowsky-Str. 1/Bau 24, 91058 Erlangen, Germany}, J. Janousek, and H.-A. Bachor}

\affiliation{{ ARC Centre of Excellence for Quantum-Atom Optics, The Australian National University, Canberra
ACT 0200, Australia\\
}}

\begin{abstract}
 Squeezing experiments which are capable of creating a minimum uncertainty state during the nonlinear process, for example optical parametric amplification, are commonly used to produce light far below the quantum noise limit. This report presents a method with which one can characterize this minimum uncertainty state and gain valuable knowledge of the experimental setup.
\end{abstract}

\maketitle

\section{Introduction}

Experiments which produce light below the quantum noise limit (QNL) already exist since the mid 80s and several different technics, such as four wave mixing \cite{slusher}, the Kerr effect \cite{Kerreffect}, second harmonic generation \cite{SHGsq} or optical parametric amplification \cite{OPA1, OPOOsq}, have been developed to produce many squeezing results. In this report we will focus on the popular method of optical parametric amplifiers (OPA), which is capable of generating high amounts of quadrature squeezing \cite{OPAbest1, OPAbest2, OPAbest3, OPAbest4}, and theoretically can produce a minimum uncertainty state during the nonlinear process \cite{Bachor2004}. However, the minimum uncertainty state can never be measured directly as there will always be loss and noise sources which will make the detected state more noisy.  A method with which one can determine this minimum uncertainty state precisely in a quick and easy way would be very useful as one could then characterize the system more closely by knowing how much squeezing is maximum possible and determining where the losses in the setup exactly are.
\\
  A common way to describe a minimum uncertainty is by referring to the variance of the quadratures of the squeezed signal. This has been analyzed theoretically many times (for example see \cite{Milburn, Bachor2004}) and we will only focus and present the most important theoretical tools for this report.  The variance of a generalized quadrature of the squeezed beam at the rotation angle $\theta$ is given by \cite{Bachor2004}

\begin{equation}
\label{Varout}
\Delta^{2}(\textbf{X}(\theta)_{MU})=\cosh(2r)-\sinh(2r)\cos(2(\theta-\theta_{s}))
\end{equation}

 with $r$ representing the degree of squeezing and $\theta_{s}$ the squeezing angle.  The amplitude of this periodic function is determined by the degree of squeezing $r$, while its minimum and maximum are governed by the angle $\theta_{s}$. This function describes a squeezed beam in its minimum uncertainty state. However, due to losses and different noise sources, one can never measure this minimum uncertainty state directly.
 In a typical squeezing experiment losses can occur at several points. It
 commences with the  optical parametric amplifier where the
squeezed light is produced and which normally has an escape efficiency below one
$\eta_{esc}<1$. Further losses can
occur at lenses $\eta_{lens}$, beam splitters $\eta_{BS}$ and finally
at the photodetector $\eta_{det}$. The total efficiency of the
system is then defined as

\begin{equation}
\eta=\eta_{esc} \cdot \eta_{lens} \cdot \eta_{BS} \cdot \eta_{det}.
\end{equation}

 The influence
of the total efficiency $\eta$ on the variance is given by
~\cite{Bachor2004}

\begin{equation}
\label{beamsplitter}
\Delta^{2}(\textbf{X}(\theta)_{Loss})=1+\eta(\Delta^{2}(\textbf{X}(\theta)_{MU})-1)
\end{equation}

with $\Delta^{2}(\textbf{X}(\theta)_{Loss})$ being the measured
variance and $\Delta^{2}(\textbf{X}(\theta)_{MU})$ the variance of
the light in the minimum uncertainty state.

\begin{figure}
\centering
\includegraphics[width=1\textwidth]{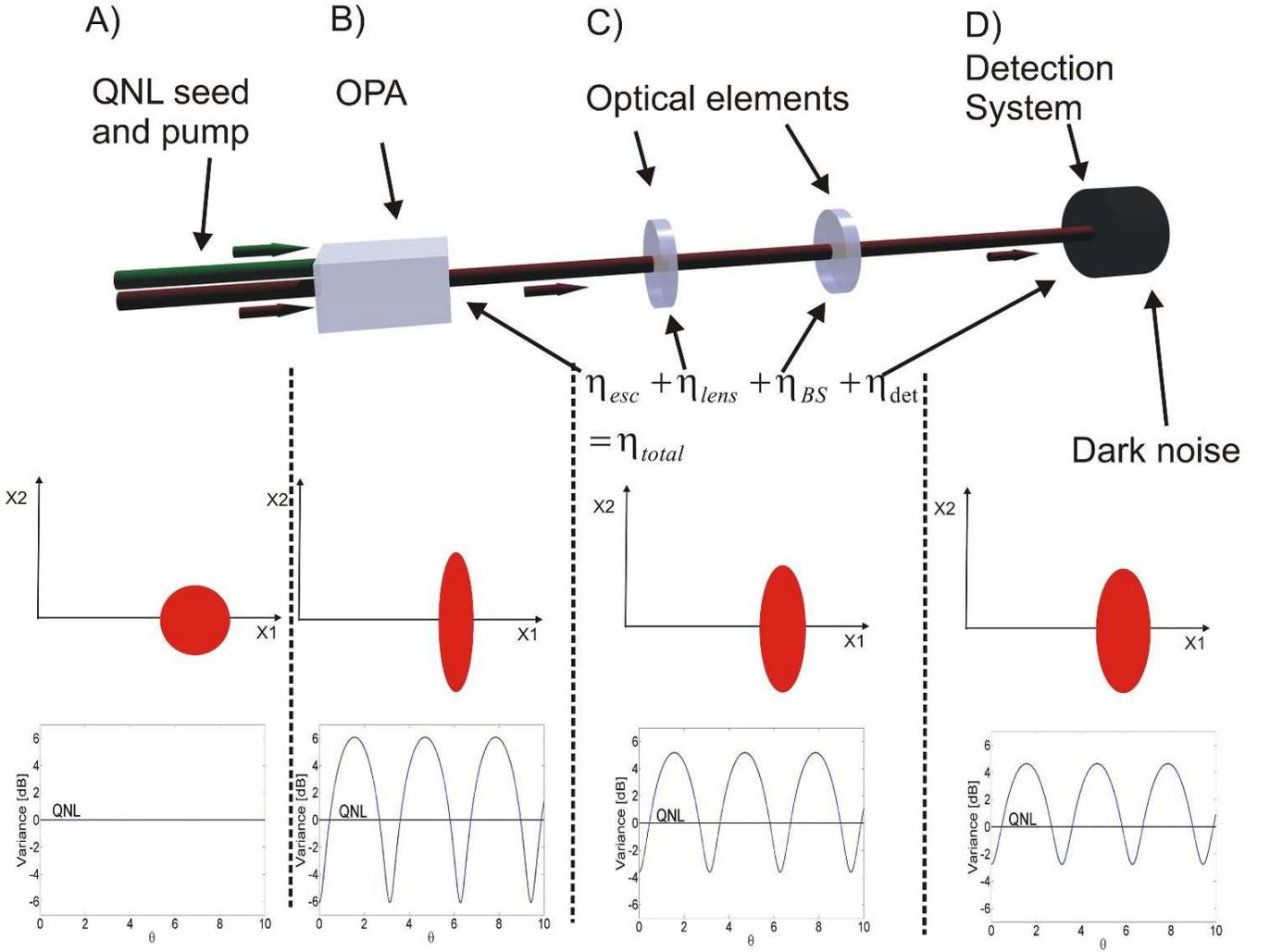}
\caption[Losses in a squeezing experiment]{A schematic of a typical squeezing experiment. Here an OPA, which is operated with a quantum noise limited pump and seed beam (part A))
 generates a squeezed beam which is in its minimum uncertainty state (part B)). It leaves the OPA which has an escape efficiency $\eta_{esc}$, goes
through optical components such as lenses $\eta_{lens}$ and beam
splitters $\eta_{BS}$  and is then detected at a detection system with a
detection efficiency $\eta_{det}$ (part C)). When the signal is processed also the dark noise from the detection system influences the signal (part D)). In this Figure the commonly used ball on a stick picture and the output variance when varying $\theta$ are used to illustrate how the minimum uncertainty becomes more noisy during the course of the experiment.} \label{lossessquezzing}
\end{figure}

Another important parameter which can influence the measured signal
is the dark noise. Even when there is no light shining on the
photodetector a small signal can still be measured. This is due to
various factors such as thermal effects, material impurities, stray
fields etc. ~\cite{Bachor2004}. In most experiments the dark noise
can be ignored as the optical power is much larger than the dark
noise. However, if one measures squeezing the dark noise can have a
big influence as only very small fluctuations are being measured.

\begin{figure}
\centering
\includegraphics[width=0.48\textwidth]{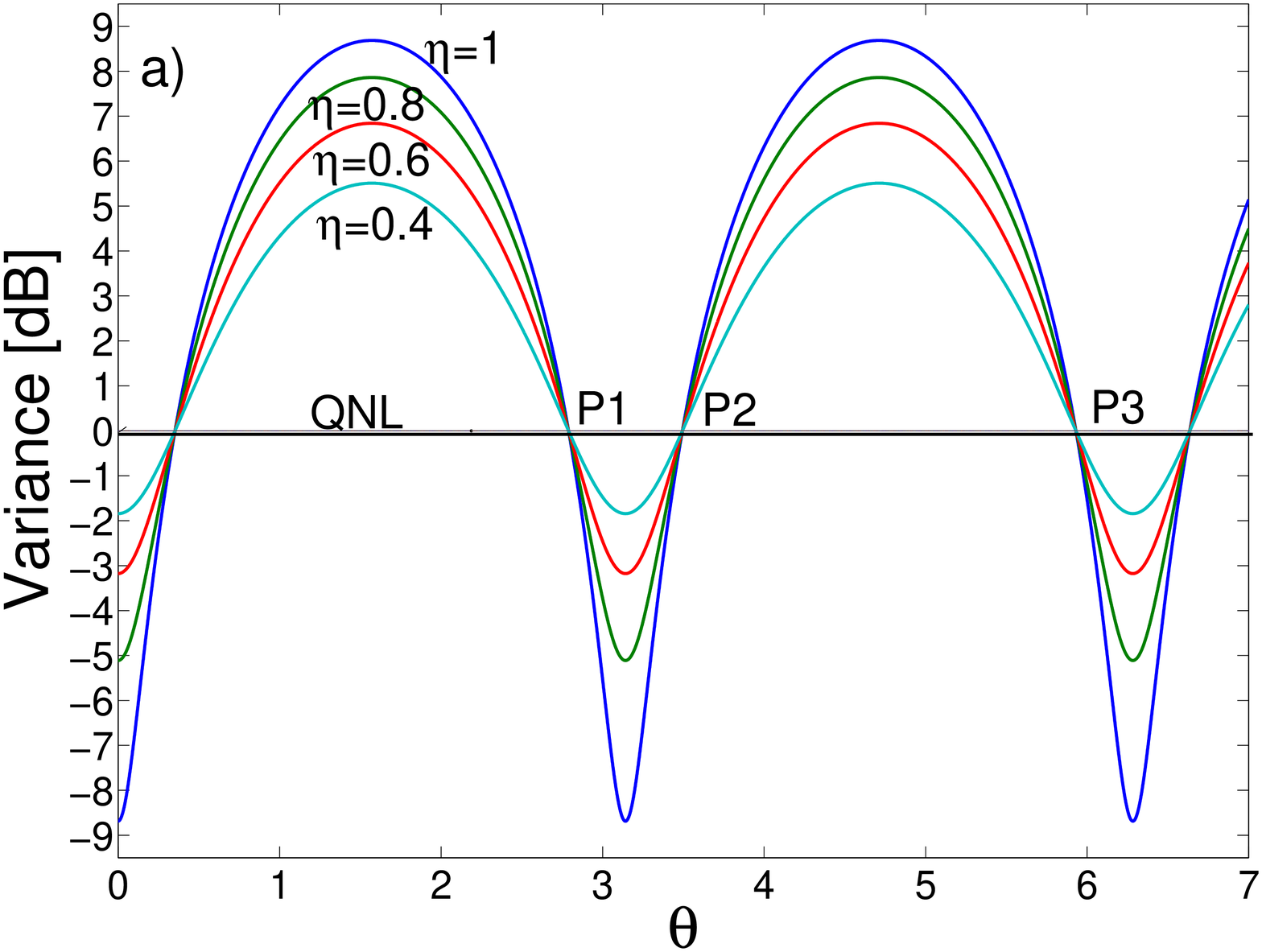}
\includegraphics[width=0.48\textwidth]{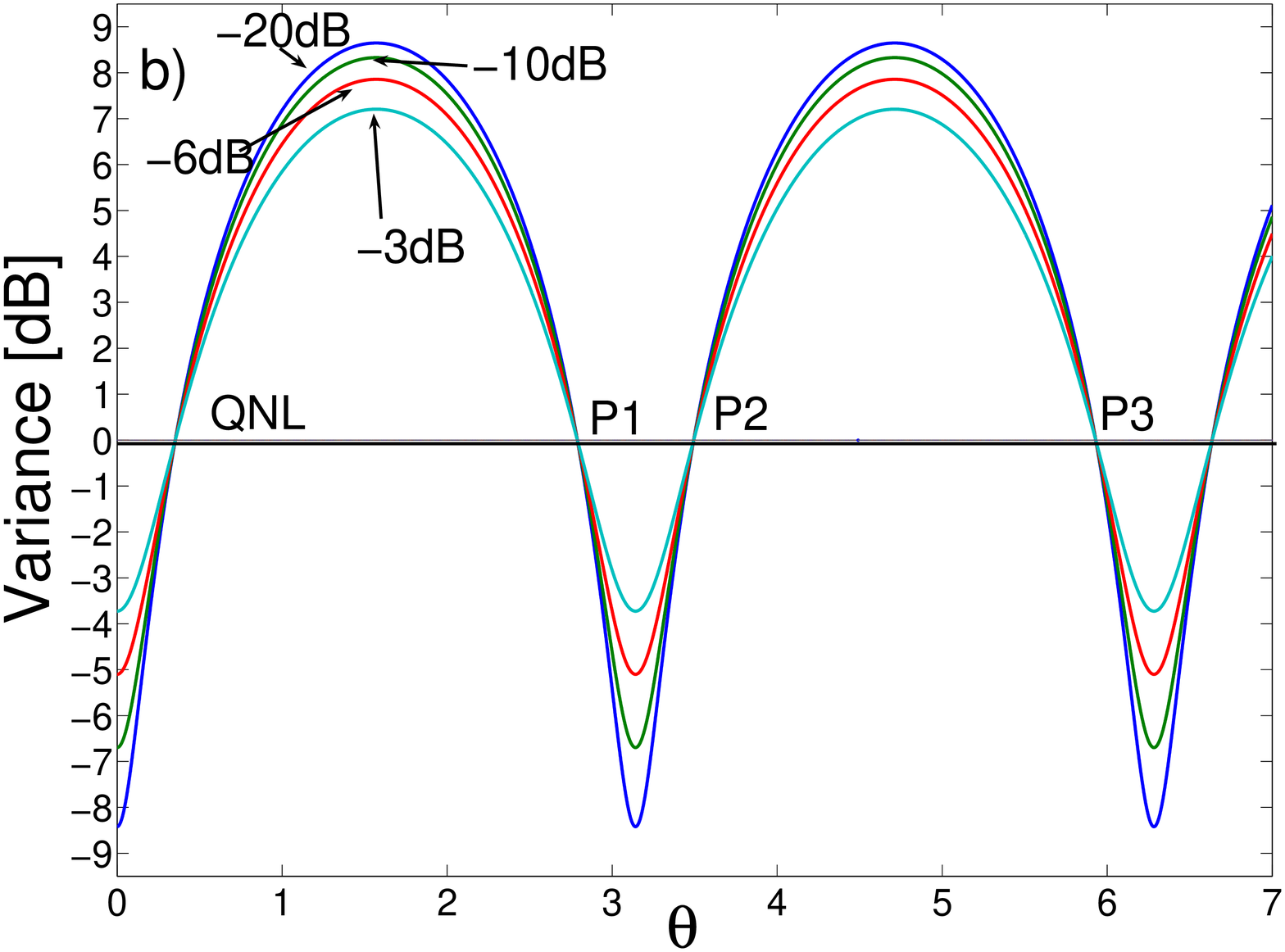}
\includegraphics[width=0.48\textwidth]{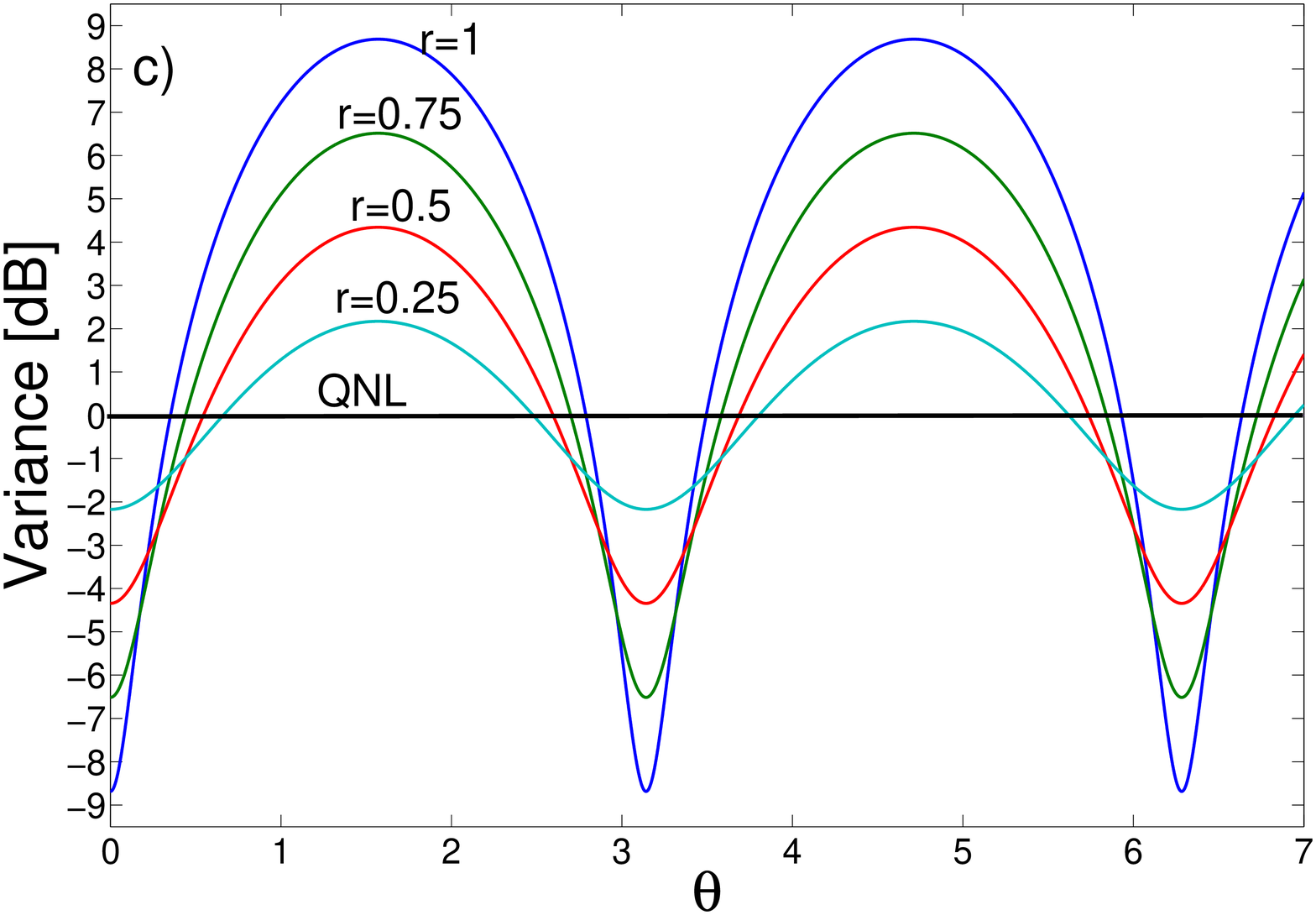}
\caption[Variance on a logarithmic plot]{Variance on a logarithmic
plot. a) $\eta$ is varied, there is no dark noise, $r=1$, b) the
dark noise is varied, $\eta=1$, $r=1$ and c) only the degree of
squeezing is varied.} \label{etas}
\end{figure}

This means that when simulating real measurements, one has to consider influences such as losses and dark noise to the variance in equation
(\ref{Varout}). The detected variance is the given by

\begin{equation}
 \Delta^{2}(\mathbf{X}(\theta)_{Det})  =  (1+\eta(\Delta^{2}(\mathbf{X}(\theta)_{MU})-1))(\Delta^{2}(X_{QNL})
-\Delta^{2}(X_{dark}))+\Delta^{2}(X_{dark})
\label{Varspec1}
\end{equation}

with $\eta$ being the efficiency, ${\Delta^{2}(X_{dark})}$ the amount of dark noise, $\Delta^{2}(X_{QNL})$ the quantum noise limit, $\Delta^{2}(\mathbf{X}(\theta)_{Det})$ the detected efficiency and $\Delta^{2}(\mathbf{X}(\theta)_{MU})$ the input state, which is assumed to be a minimum uncertainty state. It is very interesting to see what different effects, such as losses $\eta$, the dark noise ${\Delta^{2}(X_{dark})}$ and the amount of squeezing $r$, have on the output variance. This is shown in Figure
\ref{etas} a), b) and c), respectively. In the graphs the zero-line
represents the QNL. Notice that in Figure \ref{etas} a) and b) the
zero-points stay the same for all cases and only the amount of
squeezing and anti-squeezing changes. Only in Figure \ref{etas}
c), where the degree of squeezing is changed, do the zero-points
vary.

\section{Determining the Initial Degree of Squeezing}

The observation that the zero-points vary only if the degree of squeezing $r$ is changed
is very interesting. The ratio seems to be independent of the dark noise and loss and due to this might be a very helpful tool to determine the minimum uncertainty state. Let us look at three
consecutive points $P1$, $P2$ and $P3$  as marked in Figure
\ref{etas} a) or b). 
When setting $\Delta^{2}(X_{QNL})=1$, $\theta_{s}=0$ and combining it with equation (\ref{Varout}), these consecutive points are given by:

\begin{eqnarray}
x_{P1} & = & -\frac{1}{2}\arccos(l), \\
x_{P2} & = & \frac{1}{2}\arccos(l), \\
x_{P3} & = & -\frac{1}{2}\arccos(l)+\pi,
\end{eqnarray}

with $l$ being

\begin{equation}
l=\frac{-1}{sinh(2r)}+\frac{cosh(2r)}{sinh(2r)}.
\end{equation}

 This means that the only parameter that influences the zero-points is
$r$. Consequently, the ratio
$\frac{\overline{P1P2}}{\overline{P2P3}}$ only depends on $r$. It is
given by

\begin{equation}
\label{ratio}
\frac{\overline{P1P2}}{\overline{P2P3}}=\frac{\arccos(l)}{-\arccos(l)+\pi}.
\end{equation}

This is a rather powerful tool. It means that one only needs to
measure the ratio $\frac{\overline{P1P2}}{\overline{P2P3}}$ to
determine the initial degree of squeezing. The ratio can be measured
directly from the graph on the spectrum analyzer. It is not
necessary to correct the data for dark noise and similar influences
anymore. This implies that it can be a much more accurate way of
determining the minimum uncertainty state of the measured signal.
Here only the error of the ratio will influence the accuracy of
the minimum uncertainty state calculated. It should be noted again that this method is only valid if the initial state is a minimum uncertainty state.

\begin{figure}
\centering
\includegraphics[width=0.6\textwidth]{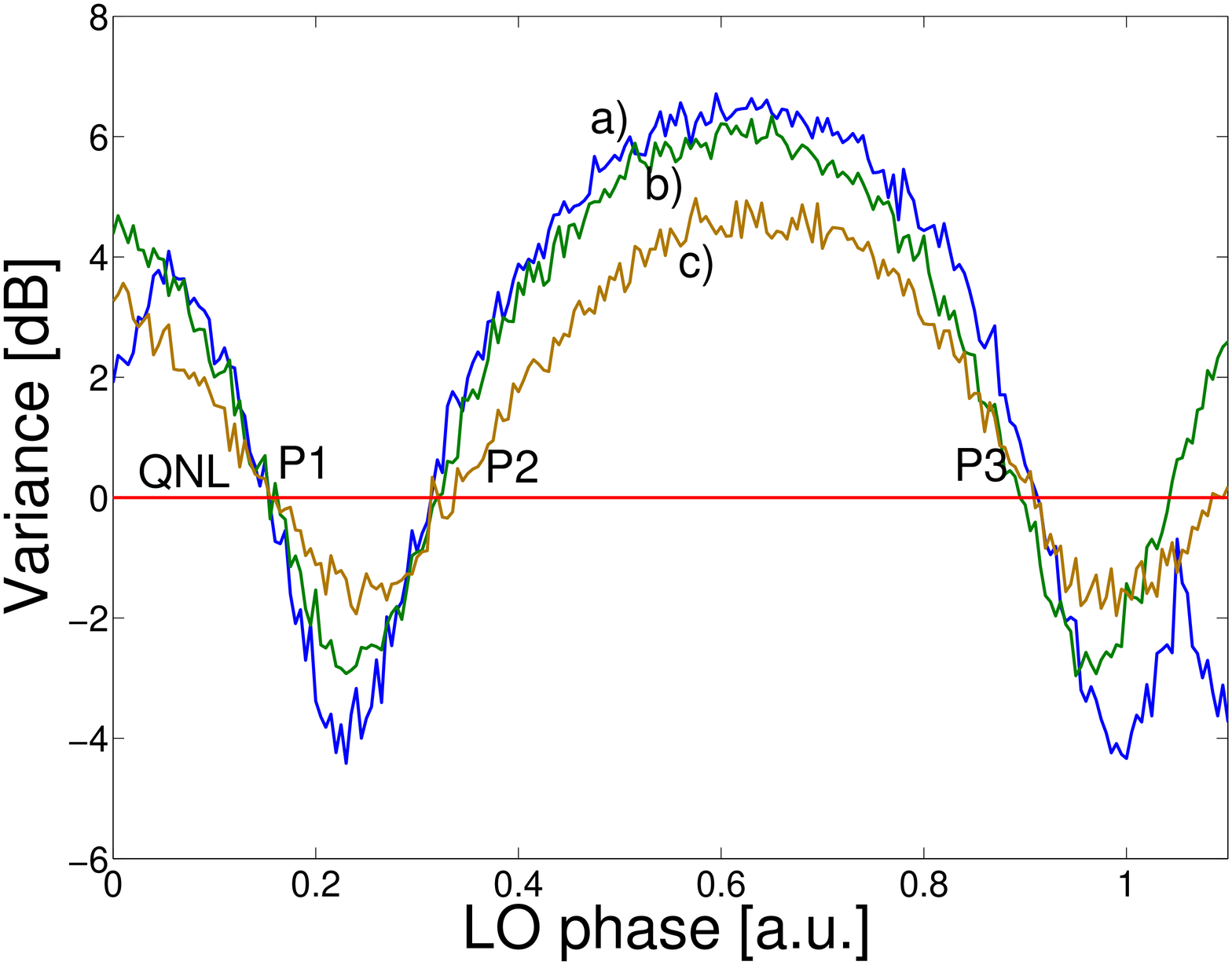}
\caption{Experimental measurements are displayed when the degree of squeezing is kept constant
while the loss is varied. Measurement a) has no attenuation while b)
and c) have 1dB and 3dB of attenuation respectively.}
\label{ratioreal}
\end{figure}

Experimentally it can be shown that losses do not influence the
ratio. This can be done simply by taking one squeezed beam and
changing the attenuation of the measured signal, which means adding
extra losses to the systems. This is displayed in Figure \ref{ratioreal} a). The OPA is
locked to de-amplification, meaning that squeezing will be measured
in the amplitude quadrature. Three different measurements were
taken: one with no attenuation, one with 1dB and one with 3dB
attenuation.  One can see that they all have
different amounts of squeezing and anti-squeezing. However, the
zero-points are all the same, meaning that the ratio is the same.
This supports the argument that only the degree of squeezing
influences the ratio and that dark noise and loss have no influence
on the ratio whatsoever. It was assumed that $\theta$, represented
here by the time axis, is on a linear scale. However, there might be
a nonlinear ramp of the PZT which can cause scaling problems such as
can be seen on the right hand side of Figure \ref{ratioreal} a).
This can also cause small errors when determining the ratio exactly.

\section{Experimental Analysis}

Let us have an even closer look at experimental data. In the experiments conducted a bow-tie cavity with a PPKTP crystal as the nonlinear medium was used as an OPA. The seed beam had a wavelength of 1064 nm and was produced inside an ultrastable,
continuous-wave single frequency laser based on Nd:YAG laser
material (Innolight GmbH, Diablo) \cite{Laser1}. The signal was detected by a homodyne detection system formed by two custom made
InGaAs homodyne detectors. The
efficiency of the detectors was $\eta_{det}=95\%$,  of the homodyne
detector $\eta_{Vis}=98\%$ and of the optical elements, such as
mirrors and lenses, $\eta_{opt}=97\%$. This gives a total efficiency
of

\begin{equation}
\eta=\eta_{esc} \eta_{det}\eta_{Vis}\eta_{opt}=\eta_{esc} \cdot 0.95\cdot 0.98 \cdot 0.97, 
\label{lossessqexperimentfinal}
\end{equation}

 with $\eta_{esc}$ being the escape efficiency of the cavity which was unknown.

\begin{figure}
 \centering
\includegraphics[width=0.6\textwidth]{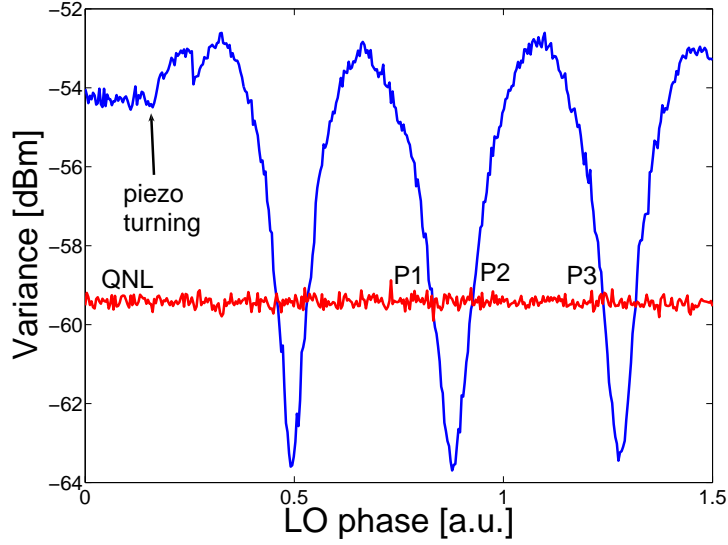}
\caption[Squeezing results]{Squeezing measurements from the bow-tie cavity. It was detected
at zero-span at 4.25 MHz with a resolution bandwidth of 300 kHz and
video bandwidth of 300 Hz. } \label{squeezingresults}
\end{figure}

The measurements were done on a spectrum-analyzer at zero-span at the detection frequency of
4.25 MHz with a resolution bandwidth of 300 kHz and video bandwidth
of 300 Hz. The OPA was locked to de-amplification, meaning that
squeezing was observed in the amplitude quadrature. One measurement can
be seen in Figure \ref{squeezingresults}. The QNL was measured to be $\Delta^{2}(X^{QNL}_{Det})=-59.4 \ \mbox{dBm}$, the maximum squeezing $\Delta^{2}(X^{+}_{Det})=-63.4 \ \mbox{dBm}$ and the anti-squeezing $\Delta^{2}(X^{-}_{Det})=-52.5 \ \mbox{dBm}$. This gives relative values of  $\Delta^{2}(X^{+}_{Det})-\Delta^{2}(X^{QNL}_{Det})=-4 \ \mbox{dB}$ and $\Delta^{2}(X^{-}_{Det})-\Delta^{2}(X^{QNL}_{Det})=6.9 \ \mbox{dB}$ of squeezing and anti-squeezing respectively.

Let us apply the method to determine the minimum uncertainty state which was presented above.
 When applying this method  it is important to note
  that the phase between the local oscillator and the signal beam must be scanned on a linear scale. The phase is scanned with a PZT located on a mirror of which the local oscillator beam is reflected. Measurements can be seen in Figure \ref{squeezingresults}. Here it can be observed when the PZT is
changing its direction and it was assumed that between two direction
changes of the PZT the scan should be fairly linear.  18 different ratios were
measured and a mean value of

\begin{equation}
\frac{\overline{P1P2}}{\overline{P2P3}}=0.307\pm0.02
\end{equation}

  calculated. From this ratio
the degree of squeezing can be determined, by using equation
(\ref{ratio}), to be

\begin{equation}
r=0.948\pm0.05.
\end{equation}

\begin{figure}
 \centering
\includegraphics[width=0.6\textwidth]{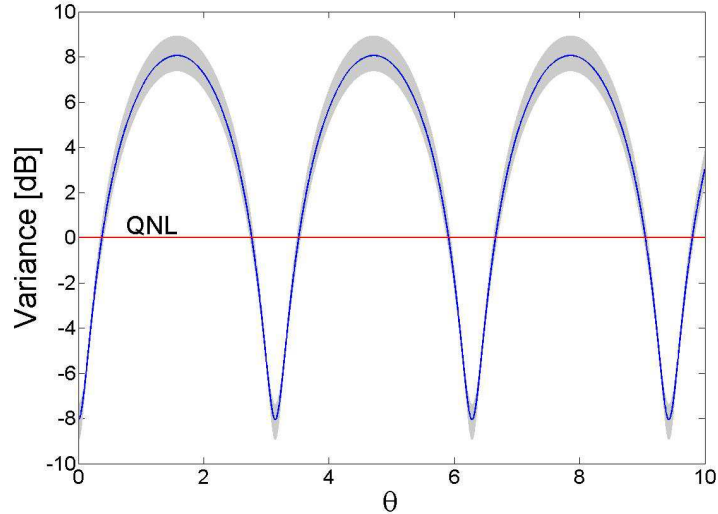}
\caption[Minimum uncertainty state inside bow-tie]{The minimum
uncertainty state which is produced inside the bow-tie cavity when calculated with the ratio-method. The grey-area represents the error margin.}
\label{minuncbow}
\end{figure}

By applying equation (\ref{Varout}) a minimum uncertainty state
with squeezing and anti-squeezing of

\begin{eqnarray}
\Delta^{2}(X^{+}_{MU})=10log_{10}(cosh(2\cdot 0.948)-sinh(2\cdot 0.948))&=&\mbox{-8.23dB} \pm0.4 \ \mbox{dBm} \\
\Delta^{2}(X^{-}_{MU})=10log_{10}(cosh(2\cdot 0.948)+sinh(2\cdot
0.948))&=&8.23\mbox{dB} \pm0.4 \ \mbox{dBm} \label{ratiosqasqfinal}
\end{eqnarray}

can be calculated. The minimum uncertainty state
is displayed in Figure \ref{minuncbow}. When looking at this minimum uncertainty state the following
important points should be noted:

\begin{itemize}

\item The results are only true for the frequency at which the signal was
detected (in this case 4.25 MHz). Normally, it can be expected that at
lower frequencies the squeezing increases, meaning that also the
minimum uncertainty state will be different.

\item It is also only valid for the pump power used in the
experiment. Different pump powers yield different gains which in
turn automatically mean different minimum uncertainty states.

\item Furthermore, the results are only true if the initial state was actually a minimum uncertainty state. If during the nonlinear process a minimum uncertainty was never created the results are obviously not accurate. A method on how to test the validity is presented below.
\end{itemize}

If the initial state was a minimum uncertainty state as the one calculated, a characterization of the loss
and noise of the system should be possible. With the help of equation (\ref{Varspec1}), one can derive the efficiency of the system to be
 
  \begin{equation}
 \eta=\frac{\Delta^{2}(\mathbf{X}(\theta)_{Det})-\Delta^{2}(X_{QNL})}{(\Delta^{2}(\mathbf{X}(\theta)_{MU})-1)( \Delta^{2}(X_{QNL})-\Delta^{2}(X_{dark}))}.
 \end{equation}

As we have measured $\Delta^{2}(\mathbf{X}(\theta)_{Det})$ and calculated $\Delta^{2}(\mathbf{X}(\theta)_{in})$ with the ratio method, we can calculate the total efficiency $\eta$. If one does this calculation twice with two distinctive values, such as squeezing and anti-squeezing, one can check if the minimum uncertainty calculated is correct (if the two calculated efficiencies are equal the minimum uncertainty state is correct). The
following values
 were measured $\Delta^{2}(\mathbf{X}^{-}_{Det})-\Delta^{2}(X^{QNL}_{Det})=6.9 \ \mbox{dB}\pm0.2 \ \mbox{dB}$, $\Delta^{2}(\mathbf{X}^{+}_{Det})-\Delta^{2}(X^{QNL}_{Det})=-4 \ \mbox{dB}\pm0.2 \ \mbox{dB}$, $\Delta^{2}(X^{QNL}_{Det})-\Delta^{2}(X^{dark}_{Det})=10.6 \ \mbox{dB} \pm0.2 \ \mbox{dB}$,  and calculated
 (with the ratio method) $\Delta^{2}(X^{-}_{MU})=8.23 \ \mbox{dB}\pm0.4 \ \mbox{dB}$, $\Delta^{2}(X^{+}_{MU})=-8.23 \ \mbox{dB} \pm0.4\ \mbox{dB}$. After converting these values onto a linear scale, one then gets the following efficiencies:

 \begin{eqnarray}
 \eta &=& 0.756\pm 0.05 \ \ \ \ \ \ \mbox{(calculated with $\Delta^{2}(X^{-}_{MU})$ and $\Delta^{2}(X^{-}_{Det})$)} \nonumber \\
  \eta &=& 0.774\pm 0.05 \ \ \ \ \ \ \mbox{(calculated with $\Delta^{2}(X^{+}_{MU})$ and $\Delta^{2}(X^{+}_{Det})$)} 
 \end{eqnarray} 

\begin{figure}
 \centering
\includegraphics[width=0.8\textwidth]{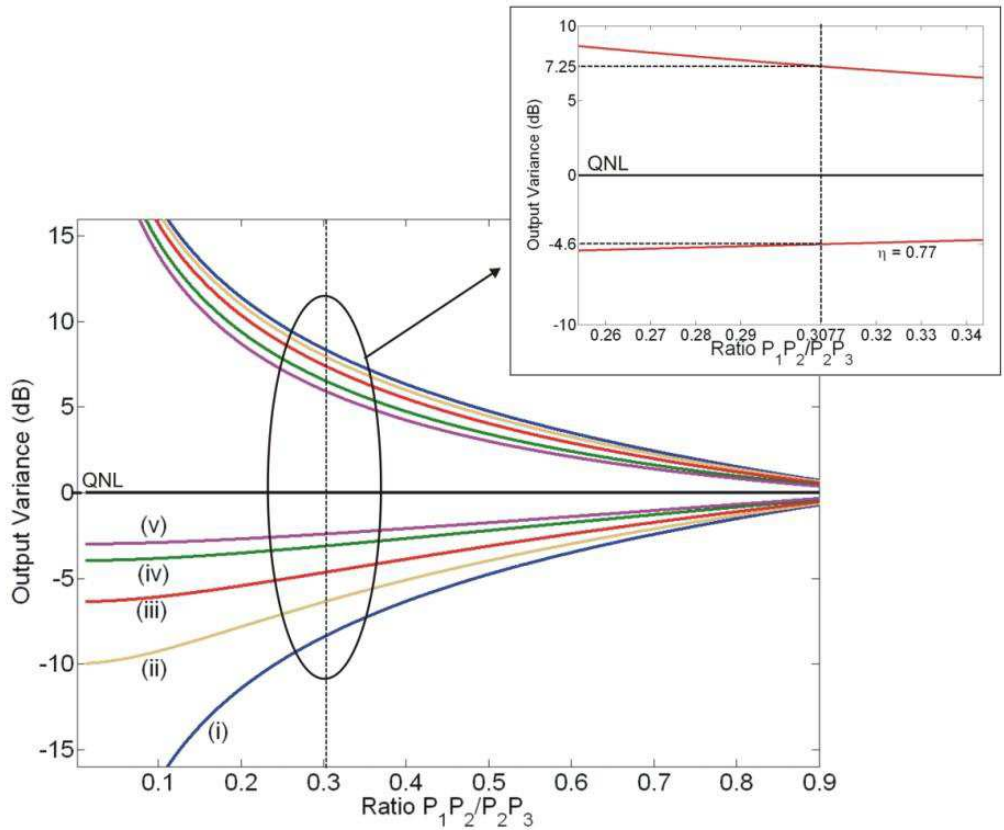}
\caption{The output variance as a function of the ratio for (i) $\eta=1$, (ii) $\eta=0.9$, (iii) $\eta=0.77$, (iv) $\eta=0.6$ and (v) (ii) $\eta=0.5$. The inset shows the case for the experimental measurements where a ratio of 0.3077 was measured.}
\label{allratio}
\end{figure}

 It can be seen that the values are nearly identical and well within the error margin.  The dependence between the output variance, the efficiency and the ratio can be seen in Figure \ref{allratio}. If there would have been a big discrepancy between these two values, this could be explained due to the fact that a minimum uncertainty state was never created, for example due to excess noise on the seed or pump beam. 

From the total efficiency calculated, we can now also determine what the escape efficiency of the OPA is. By using equation (\ref{lossessqexperimentfinal}), a value of 

\begin{equation}
\eta_{esc}=\frac{\eta}{\eta_{det}\eta_{Vis}\eta_{opt}}= \frac{0.77}{ 0.95
\cdot 0.98 \cdot 0.97}=0.85
\end{equation}
 
was calculated.

\section{Conclusion}

We can conclude that the ratio method is a quick and easy way to characterize an optical parametric amplifier. It allows us to determine what the initial minimum uncertainty state is and see how much squeezing is possible with the current experiment. From this it is also possible to quickly calculate the total loss of the system and determine other unknown variables such as the escape efficiency from the OPA. All this gained knowledge can be used to understand and improve the current setup.  This method also allows one to determine whether a minimum uncertainty state was ever created or not. In systems such as an OPA this is theoretically the case but different effects, such as noisy seed or pump beams, can prohibit it. The gained knowledge can be used to further improve and characterize the experiment.

\end{document}